\UseRawInputEncoding 
\documentclass{resonance}
\begin{document}

\title{GAIA:}
\secondTitle{The 3D Milky Way Mapper}
\author{Priya Hasan}

\maketitle
\authorIntro{\includegraphics[width=2cm]{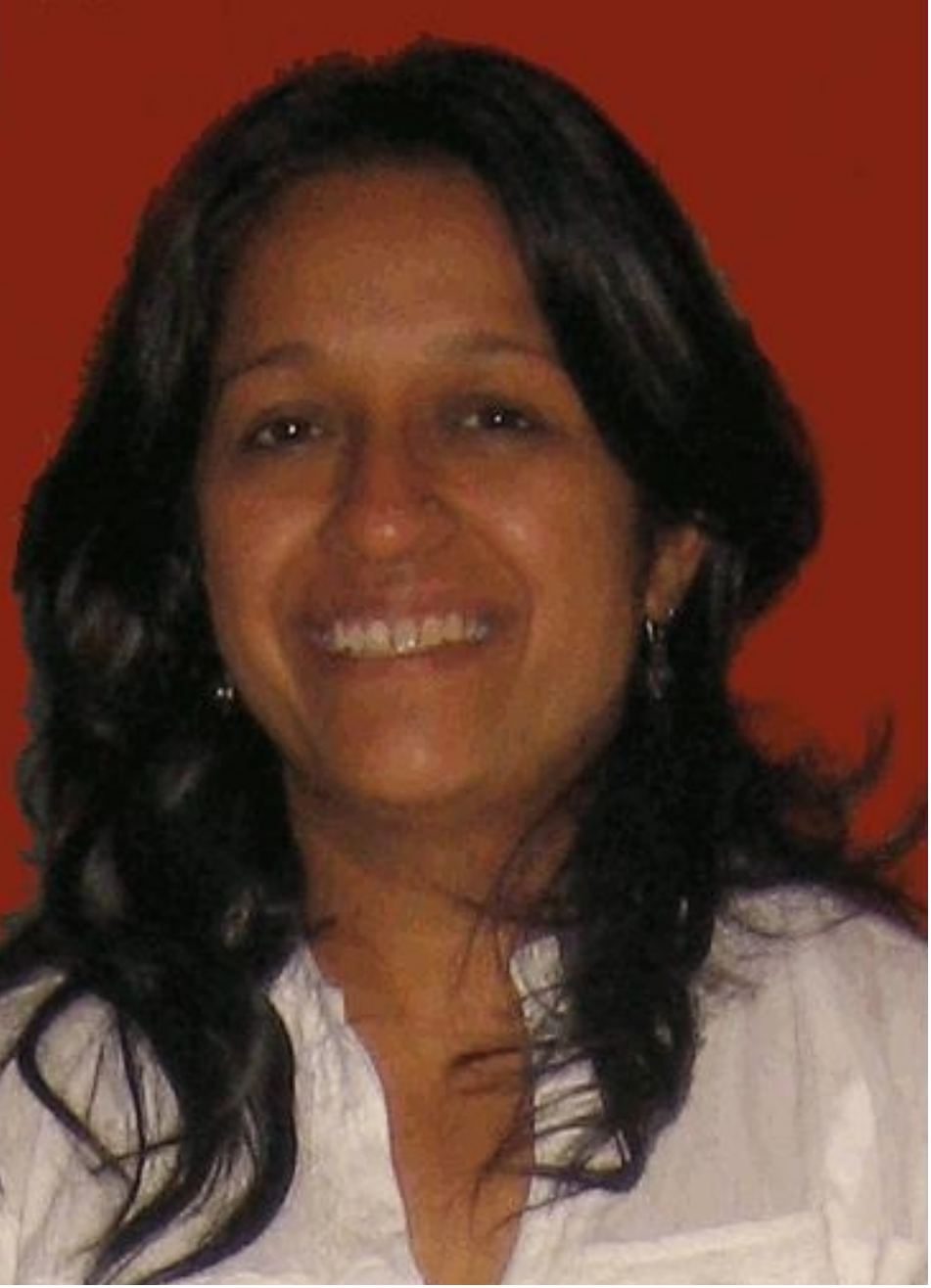}\\
Dr Priya Hasan is an Asst Professor in Physics at the Maulana Azad National Urdu University, Hyderabad. Her research interests are observational astronomy, star formation, star clusters and galaxies.  She is actively involved in olympiads, public outreach and science popularization programs. She is a member of the Astronomical Society of India - Public  Outreach \& Education Committee (since 2016)}
\begin{abstract}
GAIA (originally the acronym for Global Astrometric Interferometer for Astrophysics) is a mission of the European Space Agency (ESA) which will  make the largest, most precise three dimensional map of our Galaxy by an unparalleled survey of one per cent of the galaxy's population of 100 billion stars to the precision of micro arcseconds. This article will briefly review Gaia, the data releases and the possible implications of this mission. The reader will be introduced to  the DR1 and DR2 data releases and the scientific outcomes of DR1  as a forerunner to the much awaited DR2 of this one-of-a-kind mission.
\end{abstract}

\monthyear{April 2018}
\artNature{GENERAL  ARTICLE}

\begin{center}

\it{“And those who were seen dancing were thought to be insane by those who could not hear the music.”
  \\
  Friedrich Nietzsche}
 \end{center}

\section*{Introduction}
Imagine you are on a dance floor where everyone is dancing to their own beat, own music, own style. And what appears to be insane movements are actually very ordered movements,  to a different beat. Space is the dark labyrinth, which makes us lose our perception of depth in its two dimension projection. What we need is to hear the music and follow the dance steps of each dancer. That is exactly what Gaia is up to. 
 
On the 19th of December, 2013 Gaia was launched and put in a Lissajous-type orbit, around the Lagrange point L2 of the Sun-Earth system, which is located 1.5 million km from the Earth in the anti-Sun direction.  Ironically, Gaia (mother Earth in Greek mythology), is the offspring of Chaos, but it is expected that it will bring order at the Milky Way level to  our understanding of the positions and motions of stars. In its expected lifetime of five years, Gaia will observe each of its one billion sources about 70 times, resulting in a time-based record of the brightness and position of each source over time. The photometry and spectroscopy will also provide ages (for stars), chemical compositions and velocities of celestial objects. The scope of Gaia's potential discoveries makes the mission unique in scope and potential returns\footnote{For complete details on Gaia: http://sci.esa.int/gaia/}.

\section{Astrometry}

Astrometry \rightHighlight{Astrometry is the measurement of positions and hence motions of celestial objects. Examples of angular sizes:\\
One arcsecond ($1''$): A Rupee coin(new) viewed from a distance of 5 km.\\
One milliarcsecond (mas): An astronaut on the Moon viewed from Earth or the diameter of a human hair seen from 10 km. \\
One microarcsecond ($\mu$as): One Bohr radius ($5.29 \times 10^{-11}$m viewed from a distance of 1 m (Gaia astrometry precision). }
is the branch of astronomy which deals with the accurate measurement of the
positions and hence distances and motions of celestial objects including planets and other solar system bodies, stars within our Galaxy, galaxies and quasars.  Since observing and making catalogues of the positions of the stars and planets on the sky was important for  time-keeping, navigation and agriculture,  astronomy and astrometry were synonymous in earlier times. 
 Astrometry has applications ranging from validation of the heliocentric theory, to detection of the massive stellar black hole in the center of our galaxy, to clues to the origin of the Universe. It provides basic information related to distance and motion of celestial bodies which is essential to convert apparent measured quantities to absolute, physical ones.

\begin{figure}[!t]
\caption{Distance to a star can be calculated with simple trigonometry from the measured parallax angle (Image Credit: ESA/Medialab)}
\label{para} 
\vskip -12pt
\centering
\includegraphics[width=5.5cm, height=5.5cm]{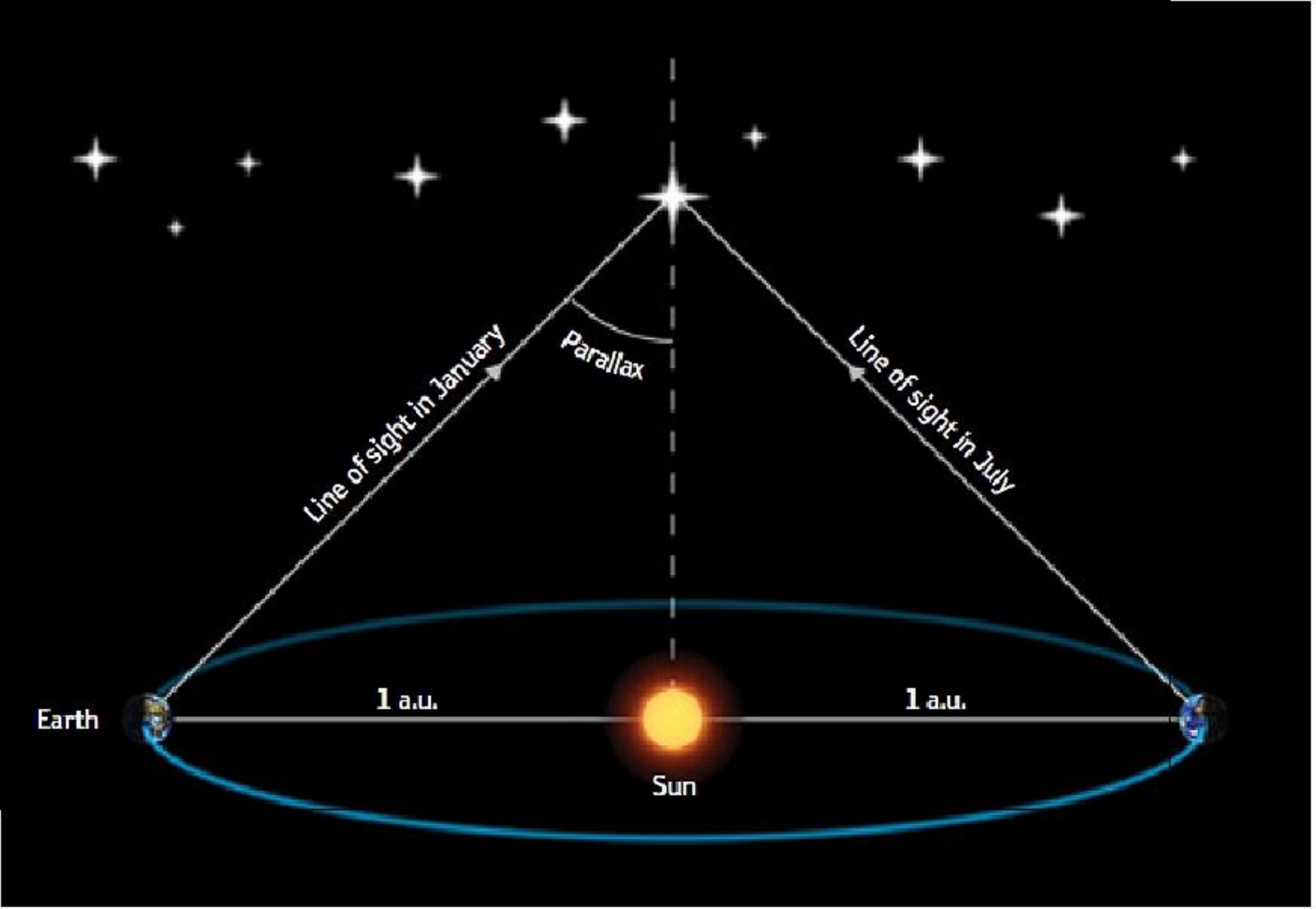}
\end{figure}

Distances to stars are really vast. If we were to scale the Sun to a marble one centimeter in size, the Earth would be a grain of salt one meter from it, Pluto would be at a distance of forty meters, and the closest stars Proxima and Alpha Centauri, would be two hundred kilometers away. The simplest method of measuring distance, is by parallax\footnote{This method is what is used by our brain to estimate distances. The distance between our eyes is a baseline and the angle estimated coupled with trignometry by the brain, leads to reasonably good distance estimates. And therefore, in case you use only one eye, it is not possible to estimate distances.}. An increase in baseline can be obtained by measuring positions of stars six months apart. By definition, a parallax ($\pi$) of one arcsecond ($1''$)  is obtained for an object at a distance of 1 pc with a baseline of 1 AU\footnote{1 AU is an astronomical unit, which is the mean distance of the Earth from the Sun or 149.6 million km.} (Fig. \ref{para}).

\section{Spacecraft}

\leftHighlight{Hipparchus (Greek astronomer, 190-120 BC), made the first catalogue of 1000 stars. He also discovered that the Earth's axis slowly changes its direction in space (precession).\\}

Since the nearest stars are at a distance much greater than 1 pc and ground-based optical telescopes have a limiting resolution due to diffraction ($0.1''$ for a 1 meter telescope in the visible), it is essential to go to space for precision astrometry\footnote{Good sites, and adaptive/active optics give significantly better than an arcsec resolution.}. Hipparcos (HIgh Precision PARrallax COllecting Satellite) was launched in 1989 to provide precision astrometry.

The Tycho Catalogue was one of the primary products of Hipparcos, which collected data for four years, from November 1989 to March 1993 for one million stars to better than 25 milliarcseconds by observing them 130 times over the duration of the mission. \leftHighlight{\textbf{Gaia: FAST FACTS}\\Launch date:19-12-2013\\
Mission end:	nominal mission end after 5 years(2019)\\
Launch vehicle:	Soyuz-Fregat\\
Launch mass:	2030 kg (710 kg of payload, 920 kg service module, 400 kg of propellant)\\
Orbit:	Lissajous-type orbit around L2\\
Partnerships:ESA} 
In addition to J2000 positions and proper motions, it included $B$ and $V$ magnitudes. Using a more advanced reduction technique, the subsequent Tycho-2 catalog was released for 2.5 million stars complete to a $V$ magnitude of 11. 

\begin{figure}[!t]
\caption{Comparison of the accuracy of astrometry measurements by previous systems and  Gaia  (Image Credit: ESA)}
\label{astrom} 
\vskip -12pt
\centering
\includegraphics[width=7.5cm, height=5.5cm]{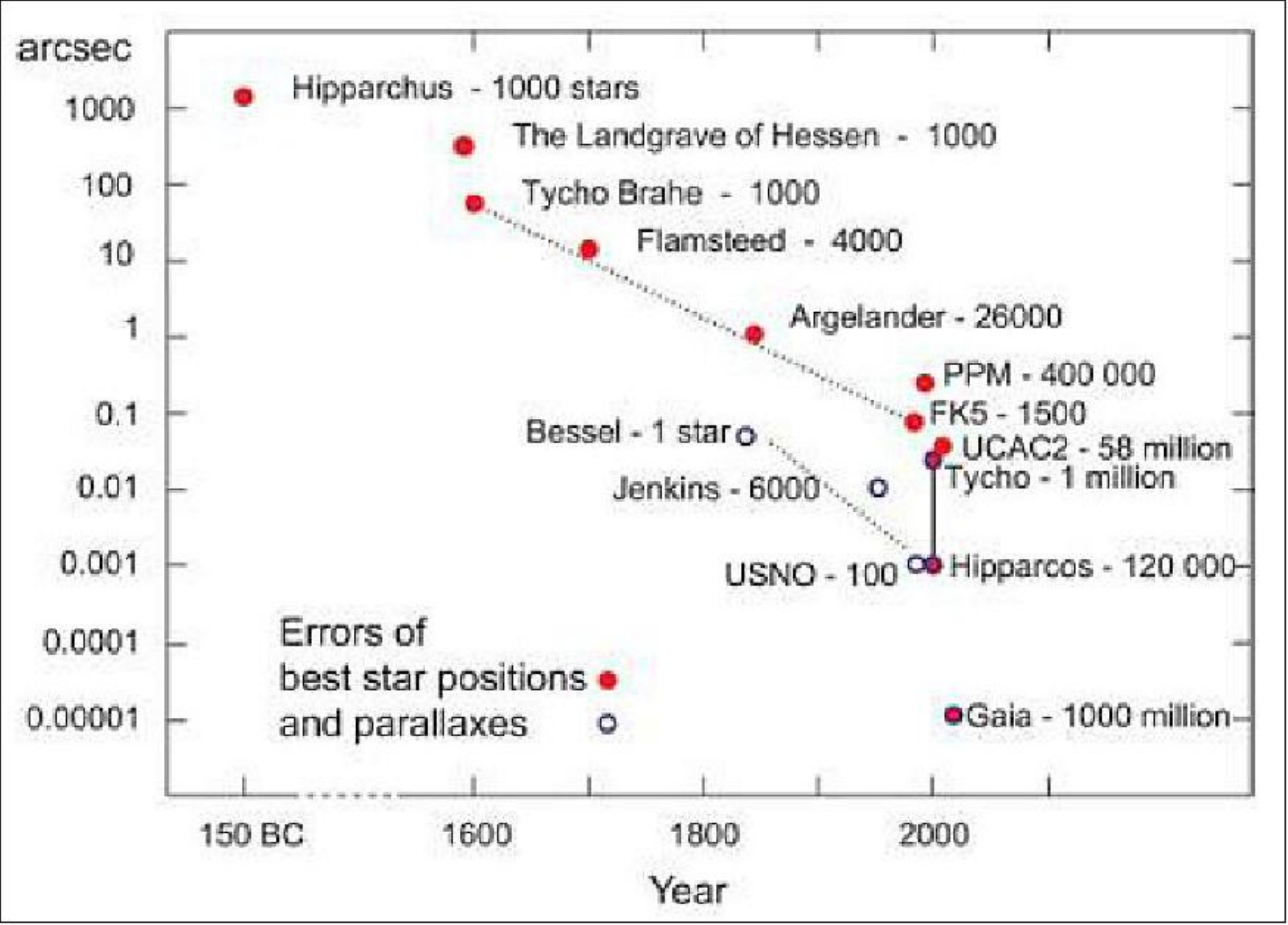}
\end{figure}

Gaia is a major improvement to the Hipparcos mission in terms of the collecting area of the primary mirrors. It collects more than 30 times the light of its predecessor and its parallax measurements are of the order of $\mu$as. To maximize collection of photons, it uses a broad $G$ band filter which centers at 5328\AA \   and with a width of 440\AA. For color information, it uses two filters $G_{BP}$ and $G_{RP}$, that center at 5320\AA \  and 7970\AA \ and with widths of 253\AA \  and 296\AA \  respectively. Figure \ref{astrom} shows the evolution of  precision astrometry in astronomical observations. 

\rightHighlight{\textbf{Gaia Instruments}\\
\textbf{Astrometry}: Two three-mirror anastigmatic (TMA) telescopes (distances and proper motions)\\
\textbf{Photometry}: Blue and Red Photometers (BP/RP) (brightness and ages of stars) \\ 
\textbf{Spectrometry}: Radial-Velocity Spectrometer (RVS)(radial velocities, chemical compositions)
}The payload module also contains all necessary electronics for managing the instrument operation and processing the raw data (Gilmore et al. 2012).

The Gaia spacecraft is composed of the payload module, the service module and the deployable Sunshield. It had a launch mass of around 2 tonnes.  The payload module is built around a toroidal-shaped optical bench (about 3 m in diameter)  which provides the structural support for the single integrated instrument that performs three functions: astrometry, photometry and spectrometry.

\begin{figure}[!t]
\caption{Gaia motions: rotation every 6 hours, precession every 63 days and revolution around the Sun with L2, every year (Image Credit: ESA)}
\label{mo} 
\vskip -12pt
\centering
\includegraphics[width=8.5cm, height=6.5cm]{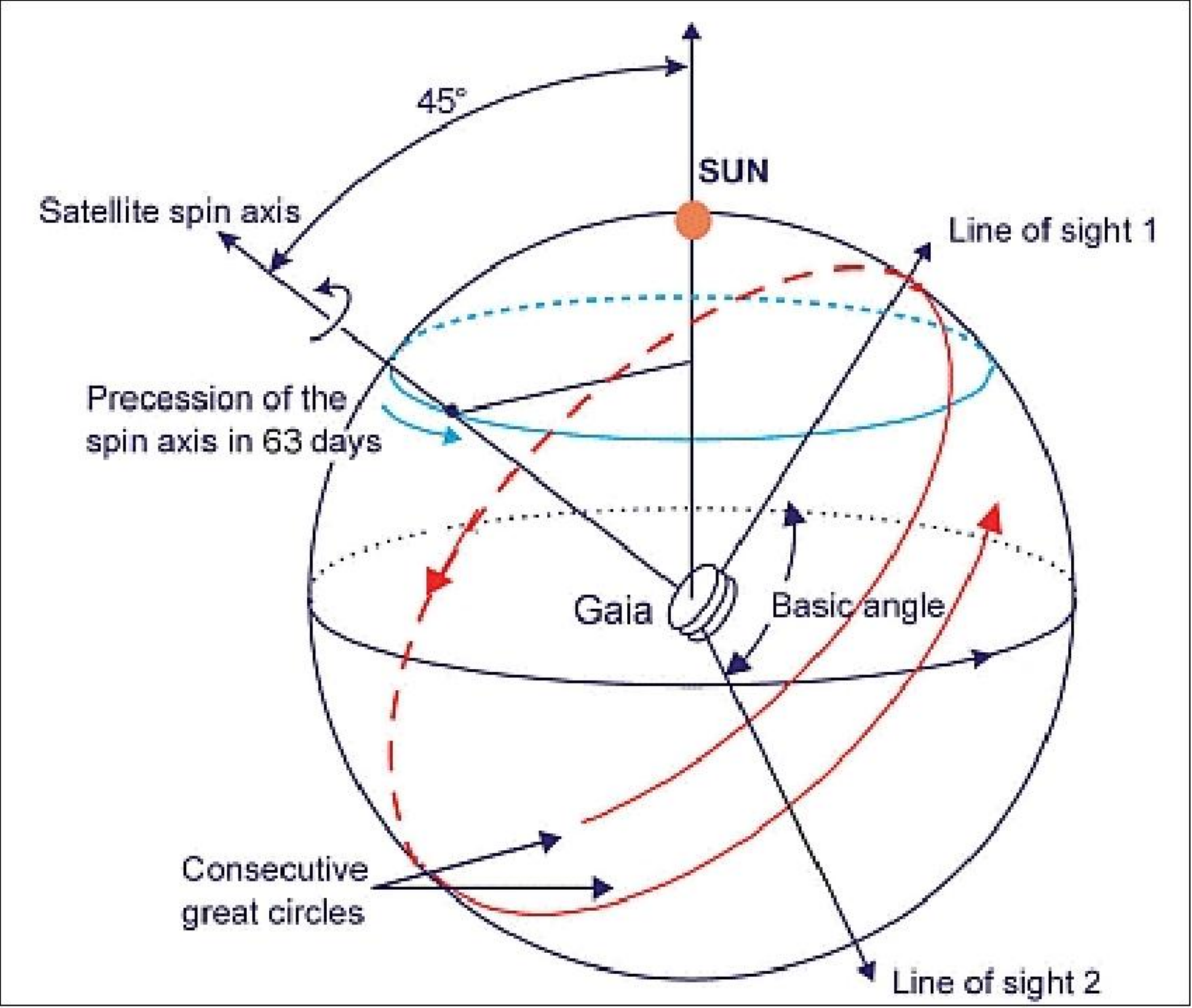}
\end{figure}

%
%
Originally Gaia was planned as an interferometric mission, but later the present optical telescope design was adopted which ensured the desired astrometry. Since observations started in July 2014, two identical TMA telescopes, with apertures of 1.45 m $\times$ 0.5 m  simultaneously observe in two regions of the sky separated by the basic angle (106.5$^o$). Gaia has three motions: rotation every 6 hours, precession every 63 days and revolution around the Sun with L2, every year (Fig. \ref{mo}).
 It observes strips of the sky as it spins and records it on 106 CCDs  detectors with a total of almost 1 billion pixels.

 With an accuracy of 24 $\mu$as, we can accurately measure a distance of $d (\mathrm {pc})=1/{\pi}(\mathrm {arcsec})=10^{6}/24$ =  42 kpc￼. So with a galactic diameter of 30 kpc and the Sun being 8 kpc from the center, we can accurately measure distances in the galaxy and its close neighbourhood. Gaia will allow astronomers to study the velocities of stars in great precision. The space velocity can be found by combining the proper motion (which can be estimated from shifts in positions of stars over long baselines, thus providing the velocity as projected on the plane of the sky) with the radial velocity, which is perpendicular to the proper motion and can be found from the Doppler shifts of spectral lines observed by  the spectrograph on board Gaia. With this outstanding data set, astronomers will trace the past trajectories of stars in the Milky Way, thus studying the dynamical history of our Galaxy. By `rewinding the track of time' in the stars' paths, we will gain new insights into the formation of our Galaxy.


 Also, with Gaia's unique combination of astrometric, photometric and spectroscopic data we can characterise stars and measure their physical parameters such as their masses and luminosities and their ages. This will be an unparalleled census of the galaxy's stellar population.

\section{Issues}
\rightHighlight{Gaia data is enormous, representing 500 million astrometric measurements per day  by a double vision instrument that is spinning and precessing. The data challenge is processing raw satellite telemetry into valuable science products managed by the DPAC  which brings together skills and expertise from over 20 countries.}

Fifty-five million stars pass Gaia's focal plane every day, leading to 500 million astrometric measurements per day. So the DR1 catalog was built from about 200 million astrometric measurements. Astronomers are challenged with dealing with a flood of data since Gaia began its work in 2013. Even after being compressed by software, the data produced by the five-year mission will eventually fill over 1.5 million CD ROMs. This data is transmitted `raw' and needs processing on Earth to turn it into a calibrated set of measurements that can be freely used by the astronomical community and citizens. The Gaia Data Processing and Analysis Consortium (DPAC) is responsible for processing the raw data, which will be published in Gaia catalogue. ESA had not only to design and build the spacecraft itself, they also had to develop new computer software that will ensure the data can be processed efficiently once it is back on Earth.

\rightHighlight{\textbf{Unforseen problems}:\\
These are contamination of the telescope optics due to ice, stray light infiltrating the focal plane, micro-clanks (small mechanical vibrations arising from thermal effects) and larger-than-expected variations in the basic angle between the two Gaia telescopes.}

During the commissioning phase a number of unforeseen problems occurred with Gaia. These include contamination of the telescope optics due to ice, stray light (due to fibres on the edge of the Sun shield) infiltrating the focal plane, micro-clanks (small mechanical vibrations arising from thermal effects) and larger-than-expected variations in the basic angle between the two Gaia telescopes. Detailed investigations by DPAC  and prime contractor Airbus Defence and Space have identified means of diminishing these issues.

Gaia should detect and measure celestial objects down to magnitude 20.5, about 650,000 times fainter than an unaided eye can see. The precision of the measurements: astrometric, photometric, and radial velocity  depends upon the type of object and its magnitude.

The measurements provided in the first Gaia Data Release (DR1) are substantially more precise than those in existing catalogues. In the final Gaia catalogue, expected in the early 2020s, brighter objects (3-13 magnitude) will have positions measured to a precision of 5 $\mu$as, parallaxes to 6.7 $\mu$as, and proper motions to 3.5 $\mu$as per year. For the faintest stars (magnitude 20.5), the equivalent numbers are several hundred $\mu$as. )

The photometry measurements in the final catalogue will be precise at the level of milli-magnitudes. For the subset of objects for which radial velocity measurements are obtained these will be measured with a precision of 15 km/s for the fainter stars and as precise as 1 km/s for the brighter stars.

The accuracy of the distances obtained by Gaia at the end of the nominal mission will range from 20\% for stars near the centre of the Galaxy, some 30,000 light-years away, to a remarkable 0.001\% for the stars nearest to our Solar System.

\begin{figure}[!t]
\caption{ Measurements  that will be possible using Gaia data. (Image Credit: ESA/Medialab)}
\label{gaiac} 
\vskip -12pt
\centering
\includegraphics[width=10 cm, height=6.5 cm]{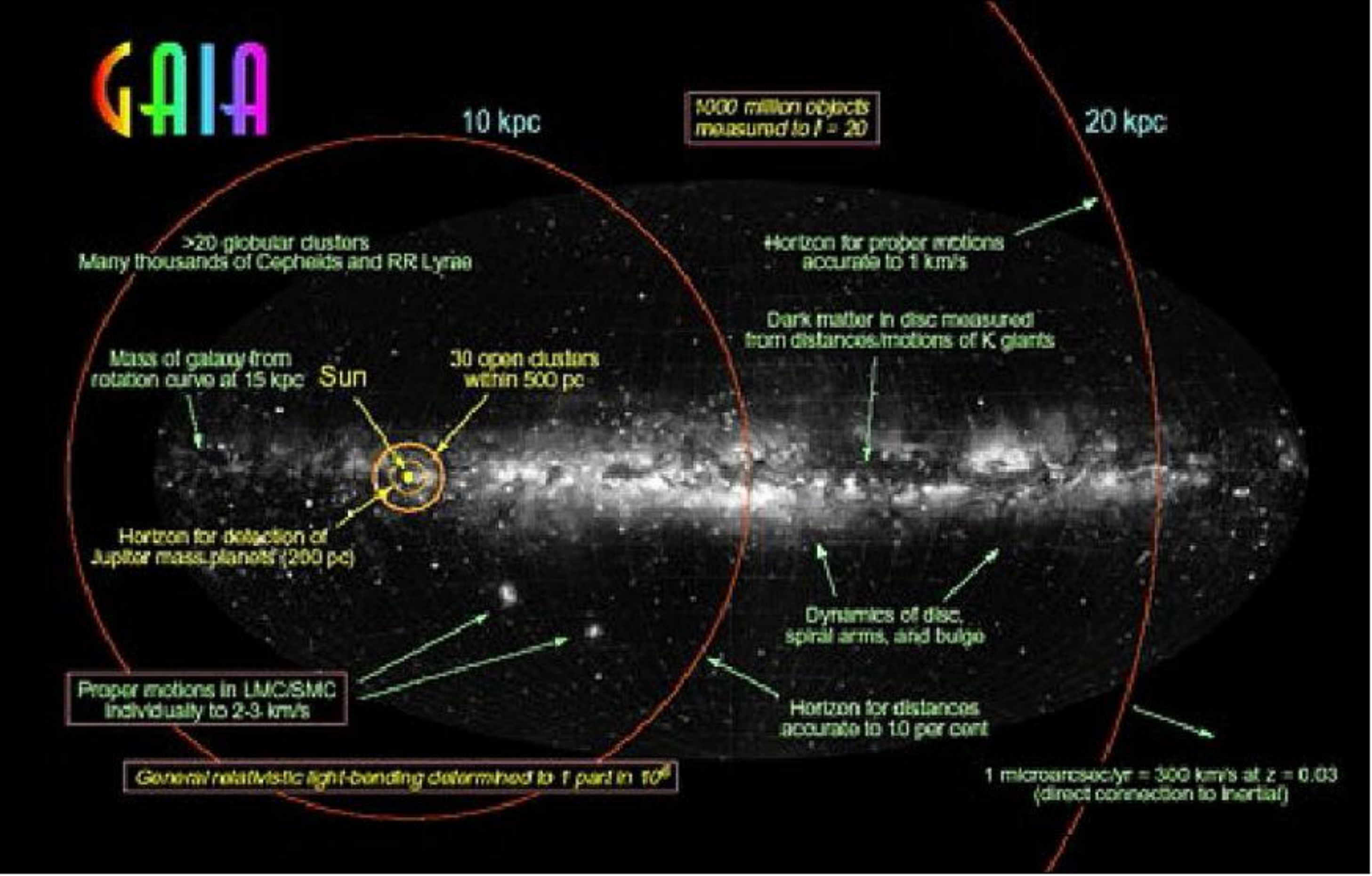}
\end{figure}

Figure \ref{gaiac} shows the kind of data we expect to get from Gaia which range from rotation curves of the galaxy at 15 kpc, proper motions in the LMC/SMC up to 2-3 km/s, the 20 kpc horizon within which we shall measure proper motions accurate to 1 km/s.


\section{Data Release 1 (DR1)}

On 14 September 2016, Gaia DR1 was released worldwide. Positions ($\alpha, \delta$) and G magnitudes for all 1 billion sources sources with acceptable formal standard errors on 
positions\footnote{See the following website for data: https://www.cosmos.esa.int/web/gaia/release}.

\begin{figure}[!t]
\caption{Gaia's first sky map. (Image Credit: ESA/Gaia/DPAC)}
\label{map} 
\vskip -12pt
\centering
\includegraphics[width=9.5 cm, height=7.5 cm]{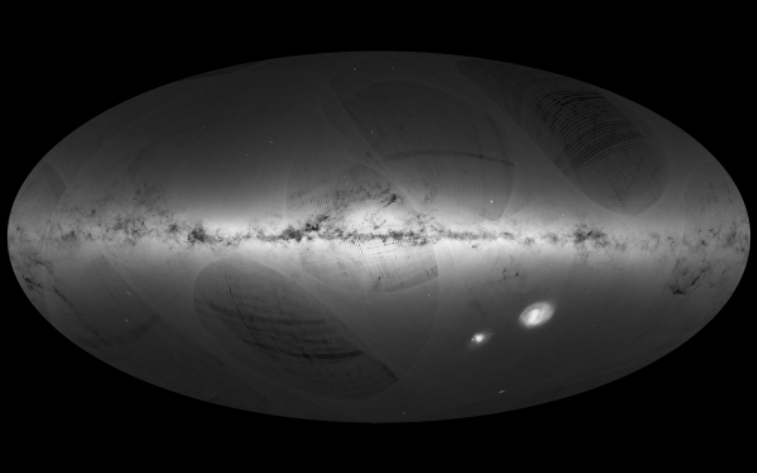}
\end{figure}

\rightHighlight{\textbf{Gaia DR1} \\
($\alpha,\delta$) for ∼ 1.1 billion sources to G = 20.7,Epoch J2015.0.\\
Five-parameter astrometric solution: positions, parallaxes, and proper motions (TGAS), brighter than 11.5 magnitude for 2 million sources.\\
Secondary sources(non-TGAS sources, only positions): 1,140,622,719\\
}

Gaia DR1 is based on observations collected between 25 July 2014 and 16 September 2015 with 5-25 transits per star through Gaia's focal plane. Figure \ref{map} shows the fist Gaia sky map released. 
 To calculate the parallax and the proper motion (along the plane of sky), it is necessary to compare observations of two diffferent epochs.  
 
 The time baseline was not sufficient for calculations of accurate parallaxes and hence distances and proper motions for DR1.  Hence, DR1 uses the data for stars common in the Tycho-2 Catalogue (measured 24 years ago) and DR1 to find the five-parameter astrometric solution: positions, parallaxes, and proper motions for 2 million stars with a precision of 0.03 mas. This part of Gaia DR1 is based on the Tycho-Gaia Astrometric Solution (TGAS). Gaia Collaboration et al. (2016, 2017) describes the DR1 in detail and its applications: performance, limitations, and future prospects.  There were various issues with DR1 due to which we have been eagerly waiting for DR2 which promises a richer treasure of data.

 In  spite of its shortcomings, there were more than 300 refereed papers from the first data release. Altmann et al. (2017) derived the movements through space of 583 million stars by comparing the Gaia positions with those of existing catalogs to obtain positions and proper motions of stars. This provided valuable data on their velocities and hence their positions in the past and future.
 \leftHighlight{\textbf{DR1 Issues\\}
Many bright stars at G $<$ 7 missing\\
Sources close to bright objects sometimes missing\\
High proper motion stars ($\mu > $ 3.5 arcsec/yr) missing\\
Extremely blue and red sources missing\\
In dense areas, effective magnitude limit may be brighter \\
Completeness problems for double stars or binaries at separations below about 4 arcsec.} 
 
Helmi et al. (2017) discovered a large population of stars that orbit the Milky Way in the opposite direction to most stars. These stars probably came from a smaller galaxy that was cannibalized by the Milky Way.
Bonaca et al. (2017) combined the data with spectroscopic surveys obtained from the ground to investigate an interesting population of stars in the halo of the Milky Way. They found that these stars have formed in-situ, rather than having been accreted from smaller galaxies, shedding light on the build-up history of our Galaxy. Oh et al. (2018) found  more than 13,000 new co-moving pairs of stars. They also found some larger associations of stars all moving in the same direction, indicating that they were all born together in the same star cluster. DR1 also provided brightness measurements of more than 3000 variable stars where about \% were newly discovered by Gaia and the positions and brightness of more than 2000 quasars, which are distant celestial objects used to define the coordinate system for astronomers to reference the sky.

\section{Data Release 2}

Gaia Data Release 2 is scheduled for 25 April 2018. The five-parameter astrometric solution for more than 1.3 billion sources, with a limiting magnitude of $G = 21$ and a bright limit of $G$ $\approx$ 3 is expected based on roughly 22 months of data and unlike TGAS, with no priors needed.

 Parallax uncertainties are in the range of up to 0.04 mas for sources at $G \leq$ 15, around 0.1 mas for sources with $G=17$ and at the faint end, the uncertainty is of the order of 0.7 mas at $G$ = 20. The corresponding uncertainties in the respective proper motion components are up to 0.06 mas yr$^{-1}$ (for $G \leq$ 15 mag), 0.2 mas yr$^{-1}$ (for $G = 17$ mag) and 1.2 mas yr$^{-1}$ (for $G = 20$ mag). The Gaia DR2 parallaxes and proper motions are based only on Gaia data; they no longer depend on the Tycho-2 Catalogue.

\rightHighlight{\textbf{Gaia DR2} \\
5-parameter astrometry for all sources \\
Broad band photometry in $G$, $G_{BP}$ , $G_{RP}$ (broad band colours), improved photometric calibrations (1 milli-mag at the bright (G$\leq$13) end to around 20 milli-mag at $G=20$), proper pass-band calibrations \\
Median radial velocities for bright ($G <$ 12), constant RV, stars.}
 
Median radial velocities will be available for more than 6 million stars with a mean $G$ magnitude between about 4 and 13 and an effective temperature in the range of about 3550 to 6900 K. This leads to a full six-parameter solution: positions and motions on the sky with parallaxes and radial velocities, all combined with mean $G$ magnitudes.
  
The photometry expected is $G$ magnitudes for more than 1.5 billion sources, with precisions varying from around 1 milli-mag at the bright ($G<$13) end to around 20 milli-mag at $G=20$. Note     that the photometric system for the $G$ band in Gaia DR2 will be different from the photometric system as used in Gaia DR1. Epoch astrometry for more than 13,000 known asteroids based on more than 1.5 million CCD observations will also be available. Light curves for more than 500,000 variable sources will also be available.

\section{Further Data Releases}

To know the 3D structure of the galaxy, we need to know the positions and space velocities of objects in the galaxy very precisely. The final Gaia catalogue will have complete astrometric, photometric, and radial-velocity catalogues as well as all available variable-star and non-single-star solutions. It will also provide color-information as well as spectroscopic data of the chemical compositions of objects, thus correlating objects with common positions, velocities, ages(for stars) and chemistry.  This will reveal important clues about the composition, formation and evolution of the Galaxy, that is, the `archaeology' of the galaxy.

It will also provide source classifications (probabilities) plus multiple astrophysical parameters (derived from photometry and astrometry) for stars, unresolved binaries, galaxies, and quasars. There will also be an exoplanet list and all epoch and transit data for all sources.

 There are various ground based spectroscopic surveys to complement Gaia data like RAdial Velocity Experiment (RAVE), APO Galactic Evolution Experiment (APOGEE),  Galactic Archaeology with HERMES(GALAH),  GAIA-ESO( European Southern Observatory) which already have some of their data available. On-ground,  proper motion surveys like  UKIRT Infrared Deep Sky Survey(UKIDSS) and Panoramic Survey Telescope and Rapid Response System (Pan-STARRS) complement Gaia data. 
\leftHighlight{\textbf{Future Data Releases:
2020:}Improved astrometry and photometry\\
Object classification and astrophysical parameters,  BP/RP spectra and/or RVS spectra \\
Mean radial velocities \\
Variable-star classifications \\
Solar-system results \\
Non-single star catalogues \\
\textbf{2022:} Full astrometric, photometric, and radial-velocity catalogues\\
All available variable-star and non-single-star solutions\\
Source classifications (probabilities) plus multiple astrophysical parameters (derived from BP/RP, RVS, and astrometry) for stars, unresolved binaries, galaxies, and quasars. \\
An exo-planet list.\\
All epoch and transit data for all sources. } 
 Various tools have been developed to aid astronomers use data like the Gaia Observation Forecast Tool (to find out when their targets will be observed by Gaia), Gaia Sky (real-time, 3D, astronomy visualisation software), Gaia-Groundbased Observational Service for Asteroids (Gaia-GOSA, to  support observers in planning photometric observations of asteroids), Gaia Ground Based Optical Tracking (GBOT, interactive tools for tracking Gaia), Gaia Universe Model Snapshot (GUMS). The reader is advised to have a look at their individual websites for further information. 
 
 Various projects will be possible  with the data at different stages. Galaxies like our Milky Way undergo dramatic dynamic changes due to gravitational interactions and merging with smaller galaxies in its neighbourhood.  Gaia will provide not only positions and motions of stars but ages and chemical compositions of stars. Thus, we can study the evolution of our galaxy over time scales of billions of years, which is also termed as `archaeology' of the galaxy.
 
  Gaia will also measure distances to extragalactic objects, exoplanets and even objects like comets and asteroids in our solar system. Stellar moving groups, clusters and streams can be identified using this information. Vital data on membership of stars in clusters and their kinematics can be well studied (Moraux 2012).  Gaia will significantly expand our knowledge of the Universe from small scales to very large scales and has a huge potential in outreach activities (O'Flaherty et al. 2008). It shall be a major repository of information for atleast the next 50 years.
  
The purpose of this article is to prepare the reader for the DR2 which is scheduled to be released on 25 April 2018. A very heavy cascade of articles based on this data is expected once this data is released. New and interesting projects can be planned to meet this excitement! All the best! Prepare and enjoy exploring Gaia!!!\footnote{A subsequent article will describe the initial results from the GAIA DR2 mission and possible projects that can be taken up.}

\section*{Acknowledgements}
The author would like to thank the referee for her/his valuable comments that helped improve the content of the article. 
\rightHighlight{\textbf{Additional Data and Tools:}\\
Spectroscopic surveys:\\
RAVE, APOGEE, GALAH, GAIA-ESO 
On-ground proper motion surveys:\\
UKIDSS,Pan-STARRS.\\ 
\textbf{Gaia tools:}\\
Gaia Observation Forecast Tool \\
Gaia Sky,Gaia-GOSA, GBOT, GUMS \\
} 



\begin{thebibliography}{99} 
\bibitem{Alt17} Altmann M., Roeser S., Demleitner M., Bastian U., Schilbach E., 2017, A\&A, 600, L4 
\bibitem{bon17} Bonaca A., Conroy C., Wetzel A., Hopkins P.~F., Kere{\v s} D., 2017, ApJ, 845, 101 
\bibitem{dr1} Gaia Collaboration et al, \textit{ A\&A}, Vol. 601, A19, 2017.
\bibitem{2016A&A...595A...1G} Gaia Collaboration et al, \textit{A\&A}, Vol. 595, A1, 2016. 
\bibitem{2012Msngr.147...25G} Gilmore G. et al., \textit{Messenger}, Vol. 147, 25, 2012.
\bibitem{Helmi17} Helmi A., Veljanoski J., Breddels M.~A., Tian H., Sales L.~V., 2017, A\&A, 598, A58 
\bibitem{mor16} Moraux, E., \textit{ EAS Publications Series},  Vol. 80-81, 73, 2016.
\bibitem{2008IAUS..248..535O} O'Flaherty K.~S., Douglas J., Prusti T.,  \textit{IAUS}, Vol. 248, 535, 2008. 
\bibitem{Oh18} Oh S., Price-Whelan A.~M., Brewer J.~M., Hogg D.~W., Spergel D.~N., Myles J., 2018, ApJ, 854, 138

\end{thebibliography}
\end{document}